\documentclass[aps,pre,showpacs,twocolumn]{revtex4}
\usepackage{amsfonts}
\usepackage[none]{hyphenat}
\usepackage[pdftex]{graphicx}

\usepackage{hyperref} 
\usepackage{cancel} 

\usepackage{appendix}
\usepackage{syntonly}
\usepackage{float} 
\usepackage{amsthm,graphics,amssymb,amsmath,latexsym}
\usepackage[all,2cell]{xy} \UseAllTwocells \SilentMatrices 
\usepackage{fancyhdr}
\newcommand{\expect}[1]{\ensuremath{\Big<#1\Big>}}

\usepackage[utf8]{inputenc} 

\begin{document}

\title{Ensemble-free configurational temperature for spin systems}
\author{G. Palma}
\email{guillermo.palma@usach.cl}
\affiliation{Departamento de Física, Universidad de Santiago de Chile,\\Casilla 307, Santiago 2, Chile.}

\author{G. Gutiérrez}
\affiliation{Grupo de NanoMateriales, Departamento de Física, Facultad de Ciencias, Universidad de Chile, Casilla 653, Santiago, Chile}

\author{S. Davis}
\affiliation{Comisión Chilena de Energía Nuclear, Casilla 188-D, Santiago, Chile}

\begin{abstract}
\vspace*{1 cm}

An estimator for the dynamical temperature in an arbitrary ensemble is derived in the framework of 
Bayesian statistical mechanics and the maximum entropy principle. We test this estimator numerically by 
a simulation of the two-dimensional XY-model in the canonical ensemble. As this model is critical in the 
whole region of temperatures below the Berezinski-Kosterlitz-Thouless critical temperature $T_{BKT}$, we use a
generalization of Wolff's uni-cluster algorithm. The numerical results allow us to confirm the robustness of 
the analytical expression for the microscopic estimator of the temperature. This microscopic estimator has also 
the advantage that it gives a direct measure of the thermalization process and can be used to compute absolute errors 
associated to statistical fluctuations. In consequence, this estimator allows for a direct, absolute and astringent 
test of the ergodicity of the underlying Markov process, which encodes the algorithm used in a numerical simulation.               
 
\end{abstract}


\maketitle

\newpage
\section{Introduction}
The concept of dynamical, or configurational temperature was made explicit for
Hamiltonian systems in the microcanonical ensemble by Rugh in 1997\cite{Rugh1997}.  
Given a particle system governed by a Hamiltonian $\mathcal{H}(\vec{q}, \vec{p})$, under the hypothesis of ergodicity, 
a microscopic functional which depends only on the position $\vec{q}$ is found to be an efficient estimator 
for the inverse temperature, $\beta=1/k_BT$. Further discussion of this idea and a generalization of the original arguments
is found in \cite{Rugh1998, Jepps2000, Rickayzen2001}. Applications and testing in molecular dynamics simulations can be 
found in \cite{Butler1998, Baranyai2000}.

An interesting generalization of the concept of dynamical temperature to classical Heisenberg spin systems 
was achieved by Nurdin and Schotte \cite{Nurdin_Schotte}. As the fundamental variables in spin systems are not 
the standard canonical conjugate $\vec{q}$ and $\vec{p}$ variables but the three components of the spin vector $\vec{S}$, which is a
constraint quantity, they used the generalized Hamilton dynamics formalism introduced in 1973 by Nambu~\cite{Nambu1973}. 
Using spin dynamics, the proposed numerical estimator for the microcanonical temperature is successfully tested in a paramagnetic spin chain. 

Further application of dynamical temperature to spin systems is reported in Ref. \cite{Nurdin2002}. In that article,
 the XY-model in one dimension (chain) as well as in a cubic fcc lattice is numerically studied by using an over-relaxation algorithm in 
the microcanonical ensemble. The microscopic estimator for the temperature gives quite reliable results and allows to perform a severe finite size analysis in the fcc-lattice close to the first order phase transition as well as in the three dimensional spin system close to the second order phase transition. They also pointed out that the estimators for temperature and other observables are not unique, which has useful and practical consequences when computing thermal averages.

In the light of the previous results, it would be desirable to have such temperature estimators for other ensembles, in addition to 
the microcanonical one. Interestingly, a generalization and extension of the concept of dynamical temperature can be obtained 
 in the framework of Bayesian statistics and the Maximum Entropy (MaxEnt) principle.  One of the more attractive features of the 
Bayesian interpretation of statistical mechanics, proposed long ago by
Jaynes~\cite{Jaynes1957}, is that it provides a general framework for setting up the probability distribution by maximizing the 
information entropy $S(F_1,F_2,..F_m)$, based on partial macroscopic knowledge 
represented by the $F_i$ quantities. This maximization of the entropy,
constrained by the given set of $F_i$, leads to the different 
probability distributions known as the different statistical ensembles.


In order to address the issue of defining an estimator for the temperature independent of the statistical ensemble, we use
the concept of conjugate variables introduced by Davis and Guti\'errez~\cite{Davis2012}.
 The main idea is to derive some general relations among expectations of microscopic functions connected with the Lagrange multipliers $\lambda_i.$
These relations are derived from the so-called conjugate variable theorem (CVT). 
Useful generalized relations for the macroscopic quantities $\lambda_i$ are obtained choosing 
suited ``trial'' microscopic functions. These microscopic quantities correspond to estimators 
of the macroscopic ones, and the obtained relations correspond to generalized hyper-virial identities.

In this paper, based on the Conjugate Variables Theorem\cite{Davis2012}, we extend the concept of dynamical temperature to 
an arbitrary ensemble, both for particle and spin systems. In the last case we
build an explicit estimator and, in the canonical ensemble, we test its performance in a Monte Carlo simulation of 
the two-dimensional XY model. The paper is organized as follows: in Sec. II an 
ensemble-independent microscopic estimator for the inverse temperature is deduced using the framework of Bayesian statistics
and the MaxEnt principle. In Sec. III the explicit analytical expression for the inverse temperature is derived for the 
two dimensional XY-model. The numerical results of the Monte Carlo simulation for this model are presented in Sec. IV, which 
include a consistency check of the statistical independence of the data obtained and a binning analysis. Finally, some essential 
consequences of having an ensemble-free microscopic estimator for the inverse temperature are discussed in Sec. V.

\section{
Temperature estimator independent of the statistical ensemble}
 
Let us consider a statistical microscopic system whose configurations are
defined by the set of $N$ variables $(x_1, x_2, ...,x_N)$, or in a compact
notation $\vec{x}$,  on a region $\Omega \in \mathbb{R}^N$. The aim of the statistical
mechanics is to find the probability distribution of the configurations $P(\vec{x})$ and the physical
properties in equilibrium of many microscopic states, compatible with a given
set of macroscopic constraints $F_1,F_2,..F_m$. As it is well known, the
solution to this problem can be expressed in terms of the maximization of the Shannon-Jaynes entropy, 
in which the constraints are included by the method of the Lagrange multipliers. The formal solution is given by the expression  

\begin{equation}
P(\vec{x}) = \frac{\exp(-\vec{\lambda} \cdot \vec{f})}{Z(\vec{\lambda})} .
\label{Prob}
\end{equation}

where $Z(\vec{\lambda})$ is the partition function defined by 

\begin{equation}
Z(\vec{\lambda}) = \int_\Omega d \vec{x} \exp (-\vec{\lambda} \cdot \vec{f}).
\label{Part_func}
\end{equation}

The vector $ \vec{f}$ is the microscopic counterpart of the macroscopic quantity
$ \vec{F}$ in the sense that its expectation value with respect to the distribution $P(\vec{x})$ is precisely $ \vec{F}$, i.e. $ \langle \vec{f}(\vec{x}) \rangle = \vec{F}$. The Lagrange multipliers are obtained implicitly through derivatives of the entropy $S$ 
\begin{equation}
\vec{\lambda} = \frac{\partial S(\vec{F})}{\partial \vec{F}}
\label{lagr_mult}
\end{equation}
where the entropy is obtained as the Legendre transform of $\ln (Z)$, $S=\ln Z+\vec{\lambda} \cdot \vec{F}$. \vspace*{0.5cm}

Now, equipped with the probability distribution given by Eq. (\ref {Prob}), the expectation value of an arbitrary scalar quantity $A(\vec{x})$ is given by the integral 
\begin{equation}
\langle A(\vec{x}) \rangle = \frac{1}{Z}\int_\Omega d \vec{x} A(\vec{x})\exp (-\vec{\lambda} \cdot \vec{f}).
\label{exp_val}
\end{equation}
By making the particular choice $A(\vec{x})= \nabla \cdot \vec{v}$ and demanding that the probability distribution vanishes on the boundary of its support, i.e. $ P(\vec{x})=0 $ for $\vec{x} \in \partial \Omega = \Sigma$, a straightforward use of the divergence theorem leads to the relation: 
\begin{equation}
 \langle \nabla \cdot \vec{v}(\vec{x}) \rangle = - \langle \vec{v} \cdot \nabla \ln P(\vec{x}) \rangle.
\label{CVT}
\end{equation}
which is called conjugate variables theorem in Ref.\cite{Davis2012}. 
Note that this identity, as written above, is not only valid for $P(\vec{x})$
given by Eq. \ref{Prob} but for an arbitrary distribution~\cite{Davis2016}. 

Now we consider the particular case in which $P$ depends 
on the configurations $\vec{x}$ through the Hamiltonian of the system $\mathcal{H}$: $P(\vec{x})=\rho(\mathcal{H}(\vec{x}))$, 
which leads to the identity: 

\begin{equation}
\langle \hspace{0.1cm} \nabla \cdot \vec{v}(\vec{x}) \hspace{0.1cm} \rangle_\rho =
\langle \hspace{0.1cm} B(\mathcal{H}   (\vec{x}) \hspace{0.1cm})  \vec{v} \cdot \nabla \mathcal{H} \hspace{0.1cm} \rangle_\rho.
\label{T_dyn}
\end{equation}
where $B(\mathcal{H})=-(d/dE) \ln \rho(E) \mid_{E=\mathcal{H}(\vec{x})} $ and
$\big<\cdot\big>_\rho$ represents an average over the ensemble characterized by $\rho$. Making the suited choice $\vec{v}=\vec{\omega}/(\vec{\omega} \cdot \nabla \mathcal{H})$, the above equation goes into
\begin{equation}
 \langle \hspace{0.1cm} B \hspace{0.1cm} \rangle_\rho = \langle \hspace{0.1cm}
\nabla \cdot \frac{\vec{\omega}}{\vec{\omega} \cdot \nabla \mathcal{H}}
\hspace{0.1cm} \rangle_\rho,
\label{T_dyn_ensemble_indep}
\end{equation}
which is the key equation for our analysis. It is worth emphasizing this
equation is independent of the particular ensemble $\rho$ used to describe the system.


In particular, if we restrict our analysis to the microcanonical ensemble, 
\begin{equation}
P_{mc}(\vec{x}) = \frac{\delta(E- \mathcal{H}(\vec{x}))}{\Omega(E)},
\end{equation}
we see that $P$ has the form $P=\rho(\mathcal{H})$ so the analysis right above
Eq. \ref{T_dyn_ensemble_indep} holds. For this case
Rickayzen\cite{Rickayzen2001}, in a generalization of Rugh's result\cite{Rugh1997}, has previously shown that 

\begin{equation}
\Big<\nabla\cdot\frac{\vec{\omega}}{\vec{\omega}\cdot \nabla \mathcal{H}}\Big>_E = \beta_{mc}(E).
\end{equation}
Note that $\beta_{mc}(E)$ is the usual definition of inverse
temperature in the microcanonical ensemble,

\begin{equation}
\beta_{\text{micro}}(E) = \frac{d}{dE}\ln \Omega(E),
\end{equation}
which is consistent with the interpretation of $\big<B\big>_\rho$ as the inverse temperature in an arbitrary ensemble.


 
For the canonical ensemble,

\begin{equation}
P_{c}(\vec{x}) = \frac{\exp(-\beta \mathcal{H}(\vec{x}))}{Z(\beta)},
\end{equation}
the probability distribution depends on the Hamiltonian as well, and therefore
the expression of Eq. (\ref{T_dyn_ensemble_indep}) holds. Now, using the particular choice 
$\vec{\omega}= \nabla \cal{H},$ one obtains an equation for the inverse temperature
as an average of the microscopic estimator in the canonical ensemble, which turns out to be
the same expression obtained in the microcanonical ensemble: 
\begin{equation}
{\beta} =  \Big \langle \hspace{0.1cm} \nabla \cdot \frac{\nabla \mathcal{H}}{\parallel \nabla \mathcal{H} \parallel^2 } \hspace{0.1cm} \Big \rangle_{\beta} \; .
\label{T_dyn_spin_CE}
\end{equation}   
%


Two comments are in order about these results. First, Eq. (\ref{T_dyn_ensemble_indep}) represents a generalization of Rugh's idea of measuring the 
temperature of a Hamilton dynamical system -restricted to the microcanonical ensemble- allowing to perform
 numerical simulations in any arbitrary statistical ensemble, 
 Secondly, Eq. (\ref{T_dyn_spin_CE}) represents, for the particular case of the canonical ensemble,
  a direct measure of the temperature. It is obtained by computing a configuration average of this estimator 
  weighted by the Gibbs factor, which contains precisely the inverse temperature. 
  In practice, one can have a computer simulation in the canonical ensemble (Monte Carlo for example),
  obtaining $\beta$ as a thermal average of the microscopic estimator

\begin{equation}
\hat{\beta} = \nabla \cdot \frac{\nabla \mathcal{H}}{\parallel \nabla \mathcal{H} \parallel^2 }
\label{estimator}
\end{equation}
  
  Moreover,  this relation allows for a direct computation of the absolute errors associated to the numerical computation of thermal averages, i.e. the efficiency of the simulation algorithm, and gives also information about the thermalization process. We 
 will illustrate these features in the case of a spin system.

\section{Inverse temperature estimator for the XY Model}
The important feature of having an ensemble-free microscopic estimators will be shown  by performing a canonical Monte Carlo simulation 
of the two-dimensional XY-model.
This model is defined by the Hamiltonian
\begin{equation}
\mathcal{H} = -J\sum_{<i,j>} \vec{S}_i \cdot \vec{S}_j,
\end{equation}
 where the angle variables $\theta_i$ describe the orientation of the unit
vectors $\vec{S}_i$ defined on a periodic square lattice of lattice size $La$
 and $J > 0$ is the ferromagnetic interaction constant between nearest neighbors denoted as $ <i,j>$. From now on we put $J=1$ and $a=1$, which sets the energy and length scales of the system.
 Our idea is to compare the  input inverse temperature $\beta_I,$ which is used as entry value in the Monte Carlo simulation, 
with the measured inverse temperature, $\beta_M,$ 
obtained as the thermal average of the microscopic estimator, given by Eq. (\ref{estimator}).

It is well known that the XY-model has a topological
phase transition at the Berezinski-Kosterlitz-Thouless temperature $T_{BKT}$
\cite{Itzykson1989, LeBellac1991}. Above this value, the relevant physical excitations are
the pairs of vortex-antivortex degrees of freedom, which destroy the quasi-order of the low temperature region, and the correlation function decays exponentially with the correlation length. Below $T_{BKT}$ the relevant degrees of freedom are the spin waves and a
Renormalization Group (RG) analysis shows that the theory is critical in the
whole range of temperature $T<T_{BKT}$, as the correlation length diverges in the
thermodynamic limit. This particular feature of the model in $d=2$, which leads
to the so-called critical slowing down effect in algorithms of local update
motivates the use of cluster algorithms as the one
implemented in the present paper (for a comprehensive discussion of this issue see Ref. \cite{Binney1992}). Nevertheless, 
as cluster algorithms generally lose their
efficiency at very low temperatures, other algorithms like over relaxation-MC
should be used~\cite{Palma2016}. Thus, this model is a demanding test for our purpose to
check that the microscopic estimator works.

In order to measure inverse temperature, we need to express
 the Rugh's estimator for the inverse temperature, Eq.(\ref{estimator}),
%
in terms of the spin variables $\vec{S}_i$.
In the case of the two-dimensional XY-model, each spin is constrained to move
in a circle, so that the full state of the system can be expressed in terms of a
vector of $N$ planar angles $\theta=(\theta_1, \ldots, \theta_N)$. The
Hamiltonian written in terms of these angles has the form
\begin{equation}
\mathcal{H}(\theta) = -J\sum_{<i,j>}\cos (\theta_i-\theta_j),
\end{equation}
and for this Hamiltonian the computation of Eq. (\ref{estimator})
 is straightforward. Moreover, we will avoid the use of the Nurdin estimator  \cite{Nurdin_Schotte}, which is written in terms of derivatives of the Cartesian
spin coordinates and involves the differential operator $\vec S \times \nabla$
in order to implement the geometric constraints.

An explicit computation of the derivatives appearing in Eq.
 (\ref{estimator}) yields

\begin{equation}
\frac{\partial \mathcal{H}}{\partial \theta_i} = J\sum_{<j\neq i>} \sin (\theta_i-\theta_j),
\end{equation}
for the gradient of the Hamiltonian, and 

\begin{widetext}
\begin{equation}
\frac{\partial^2 \mathcal{H}}{\partial \theta_i \partial \theta_j} = 
\begin{cases}
  J\sum_{<k\neq i>}\cos(\theta_i-\theta_k) & \text{if}\ i = j, \\
  -J\cos(\theta_i-\theta_j) & \text{if}\; $i$ \;\text{and}\;$j$ \;\text{are nearest neighbors}, \\
  0            & \text{otherwise,}
\end{cases}
\end{equation}
\end{widetext}
for the Hessian matrix of the Hamiltonian. We finally obtain for the estimator of $\beta$,

\begin{equation}
\hat{\beta}(\theta) = \frac{1}{|\nabla \mathcal{H}|^2}\left[\sum_i \frac{\partial^2
\mathcal{H}}{\partial \theta_i^2} - \frac{2}{|\nabla
\mathcal{H}|^2}\sum_{i,j} \left(\frac{\partial \mathcal{H}}{\partial
\theta_i}\frac{\partial \mathcal{H}}{\partial \theta_j}\frac{\partial^2
\mathcal{H}}{\partial \theta_i \partial \theta_j}\right)\right],
\label{eq_2dxy_estimator}
\end{equation}
which satisfies $\Big<\hat{\beta}(\theta)\Big>_\beta=\beta$. By introducing the notation 

\begin{equation}
g_i =  \frac{\partial \mathcal{H}}{\partial \theta_i}, \hspace{1cm} h_{ij} =\frac{\partial^2 \mathcal{H}}{\partial \theta_i \partial \theta_j}, \hspace{1cm} G = \sum_i g_i^2 
\end{equation}
we can write the microscopic estimator in Eq. \ref{eq_2dxy_estimator} in a form more suitable for direct implementation in computer
code, as follows

\begin{equation}
\hat{\beta} = \frac{1}{G}\left(\sum_i h_{ii} - \frac{2\sum_{i,j}g_i g_j h_{ij}}{G}\right).
\label{beta_estimator}
\end{equation}

\section{Results and consistency tests}

We perform a canonical Monte Carlo simulation with the Wolff uni-cluster algorithm~\cite{Wolff1989} 
for several values of $\beta_I$, corresponding to temperature $T$ between
0.1 and 2.5, with $n$=1$\times$10$^7$ Monte Carlo steps each. We have measured the inverse temperature by
using the corresponding estimator $\hat{\beta}$ given by Eq. \ref{beta_estimator}. 
The errors were estimated by using its standard deviation. 

\subsection{Performance of the inverse temperature estimator}

\begin{figure}[h]
\begin{center}
\includegraphics[scale=0.50]{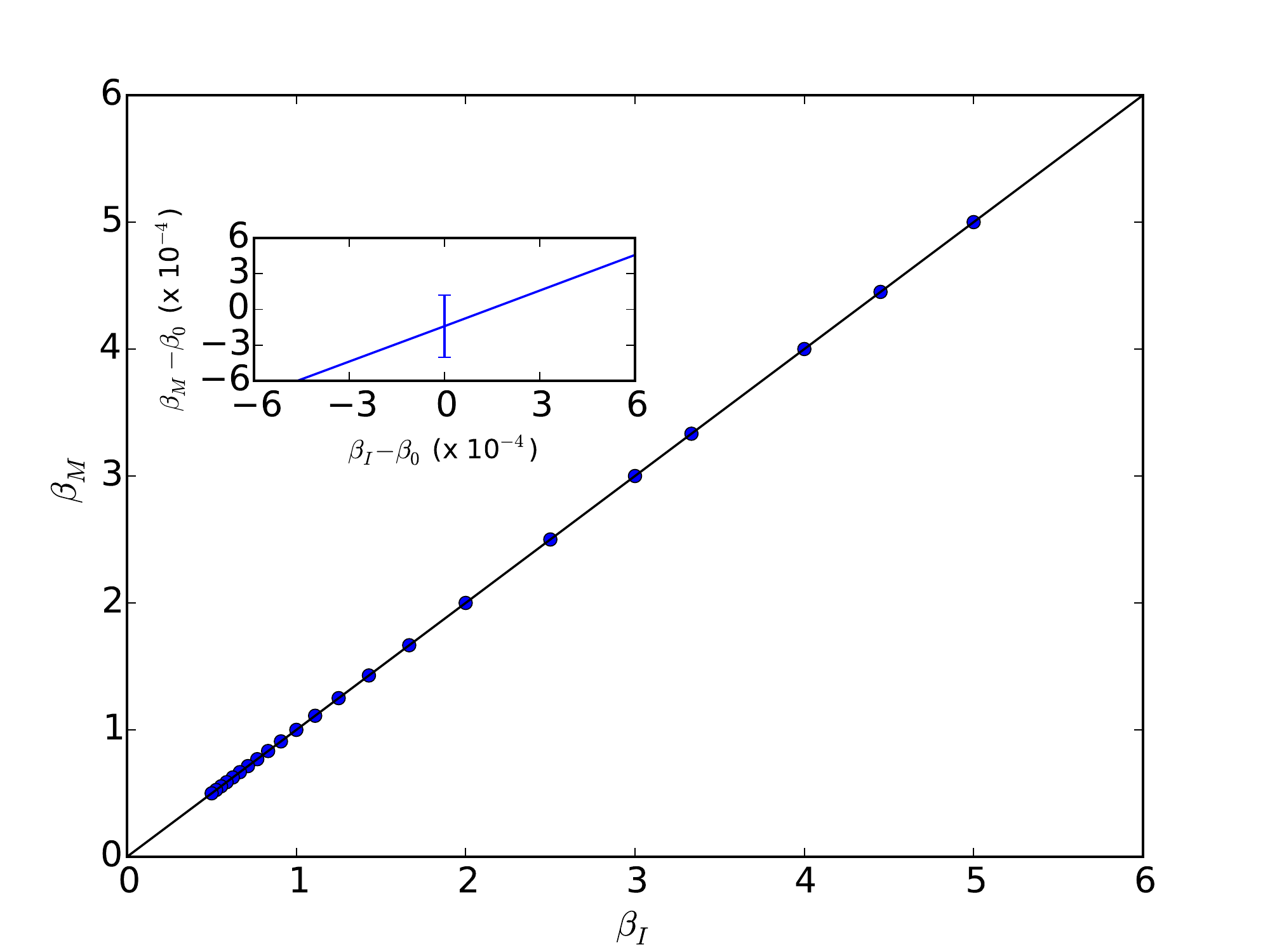}
\end{center}
\caption{Measured value $\beta_M$ as a function of $\beta_I$. The inset shows a
typical error bar, of order $10^{-4}$, which is not observable in the main plot.}
\label{fig_errorbars}
\end{figure}

The input inverse temperature $\beta_I$ and the average of the estimator $\hat{\beta}$, which is the measured 
inverse temperature $\beta_M$ are shown in Table \ref{tbl_imposed_vs_measured}, 
together with the absolute error. It can be observed that they agree up to an error less than 0.07 \%. 
%
\begin{table}[h!]
\begin{tabular}{|c|c|c|}
\hline
$\beta_I$ & $\beta_M=\big<\hat{\beta}\big>$ & Absolute error (\%)\\
\hline
10.000000 & 10.001192 & 0.007395 \\
5.000000 & 5.000141 & 0.005751 \\
4.450002 & 4.450225 & 0.005660 \\
4.000000 & 4.000208 & 0.005387 \\
3.333333 & 3.333091 & 0.005116 \\
3.000030 & 2.999870 & 0.005054 \\
2.500000 & 2.499854 & 0.004678 \\
2.000000 & 1.999990 & 0.004396 \\
1.666667 & 1.666558 & 0.004485 \\
1.428571 & 1.428484 & 0.004429 \\
1.250000 & 1.250058 & 0.004150 \\
1.111111 & 1.111117 & 0.004567 \\
1.000000 & 1.000063 & 0.005505 \\
0.909091 & 0.909065 & 0.011031 \\
0.833333 & 0.833320 & 0.019836 \\
0.769231 & 0.769018 & 0.023217 \\
0.714286 & 0.714686 & 0.028197 \\
0.666667 & 0.666893 & 0.032786 \\
0.625000 & 0.625210 & 0.037673 \\
0.588235 & 0.587817 & 0.042746 \\
0.555556 & 0.555349 & 0.046284 \\
0.526316 & 0.526175 & 0.049962 \\
0.500000 & 0.500110 & 0.050039 \\
0.476190 & 0.476198 & 0.054856 \\
0.454545 & 0.454362 & 0.058008 \\
0.434783 & 0.435149 & 0.059603 \\
0.416667 & 0.416707 & 0.069333 \\
0.400000 & 0.399700 & 0.066258 \\
\hline
\end{tabular}
\caption{Comparison of input values of temperature and averages of the estimator
$\hat{\beta}$ for $n$=1$\times$10$^7$ Monte Carlo steps.}
\label{tbl_imposed_vs_measured}
\end{table} 
The plot of Fig. \ref{fig_errorbars} shows the measured values of $\beta_M$
given by Eq. \ref{T_dyn_spin_CE} for each input value $\beta_I$ used in the
simulations, as well as their standard deviations, which are given by the expression 

\begin{equation}
\Delta \beta = \sqrt{\frac{1}{n-1}\sum_{i=1}^n \left(\hat{\beta_i}-\Big<\hat{\beta}\Big>\right)^2}.
\end{equation}
%
%

The remarkable agreement between $\beta_I$ and the average of its microscopic
estimator lets us conclude that the microscopic estimator $\hat{\beta}$ is
indeed a trustable and robust quantity to check whether the thermal averages
indeed correspond to the equilibrium values of the corresponding observables. 

\subsection{Evolution towards thermal equilibrium}

\begin{figure}[h!]
\begin{center}
\includegraphics[scale=0.5]{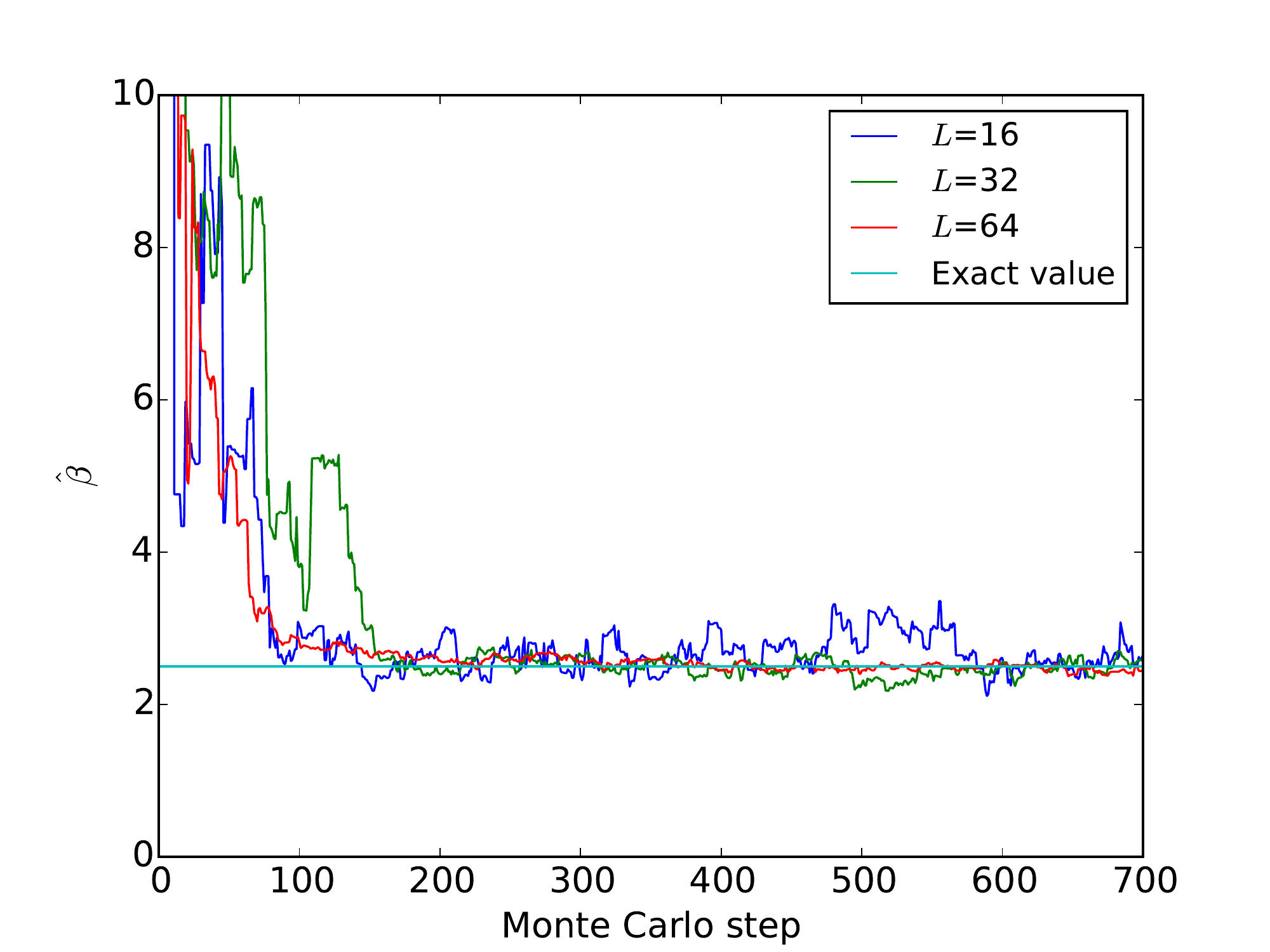}
\end{center}
\caption{(Color online) Thermalization path of the dynamical temperature estimator
$\hat{\beta}$ for system sizes of $L$=16, $L$=32 and $L$=64 at $T$=0.4.}
\label{fig_equilib}
\end{figure}

An advantage of our approach is that the estimator for $\beta$, given
by Eq. \ref{beta_estimator}, can be used to monitor the stochastic evolution towards
equilibrium of the system. In Fig. \ref{fig_equilib} we show a typical
thermalization process for systems of size $L$=16, $L$=32 and $L$=64 at a temperature $T$=0.4.
We can see that the average of our estimator yields the correct inverse temperature 
associated to the equilibrium system, which corresponds to the input value $\beta_I$. 
In all cases equilibration occurs quickly, well within 500
Monte Carlo steps. It also holds that the larger the system, the smaller the
fluctuation, as one would expect from finite-size scaling arguments. As Fig. \ref{fig_equilib} shows, 
the thermalization process turned out to be faster for larger systems at $T=0.4$, in spite of the general
statement that larger systems require a larger number of thermalization
sweeps~\cite{LeBellac2004,Janke2008}.

The instantaneous value at every Monte Carlo step
could be interpreted as the evolution of the system towards equilibrium. This is an interesting 
feature because, in a standard simulation, even if the average of some observable reaches a stationary regime, it does not 
necessarily correspond to the true equilibrium average. This may occur, for instance, in metastable systems,
such as non-extensive systems~\cite{Pluchino2004}. 
Our estimator provides an astringent test that the simulated system has thermalized, in 
the sense that the averages are compatible with the ones computed using the Gibbs distribution.




%

\subsection{Statistical independence and consistency checks}


Due to the fact that the 2$d$ XY-model is critical in the whole region below
$T_{BKT}$, i.e. it has infinite correlation length in the thermodynamic limit,
we have used a Wolff uni-cluster algorithm aiming to reduce critical slowing
down. In order to ensure the statistical independence of the generated
configurations, we have implemented different tests of consistency. 
 
\subsubsection{Autocorrelation} 

\begin{figure}[h!]
\includegraphics[scale=0.45]{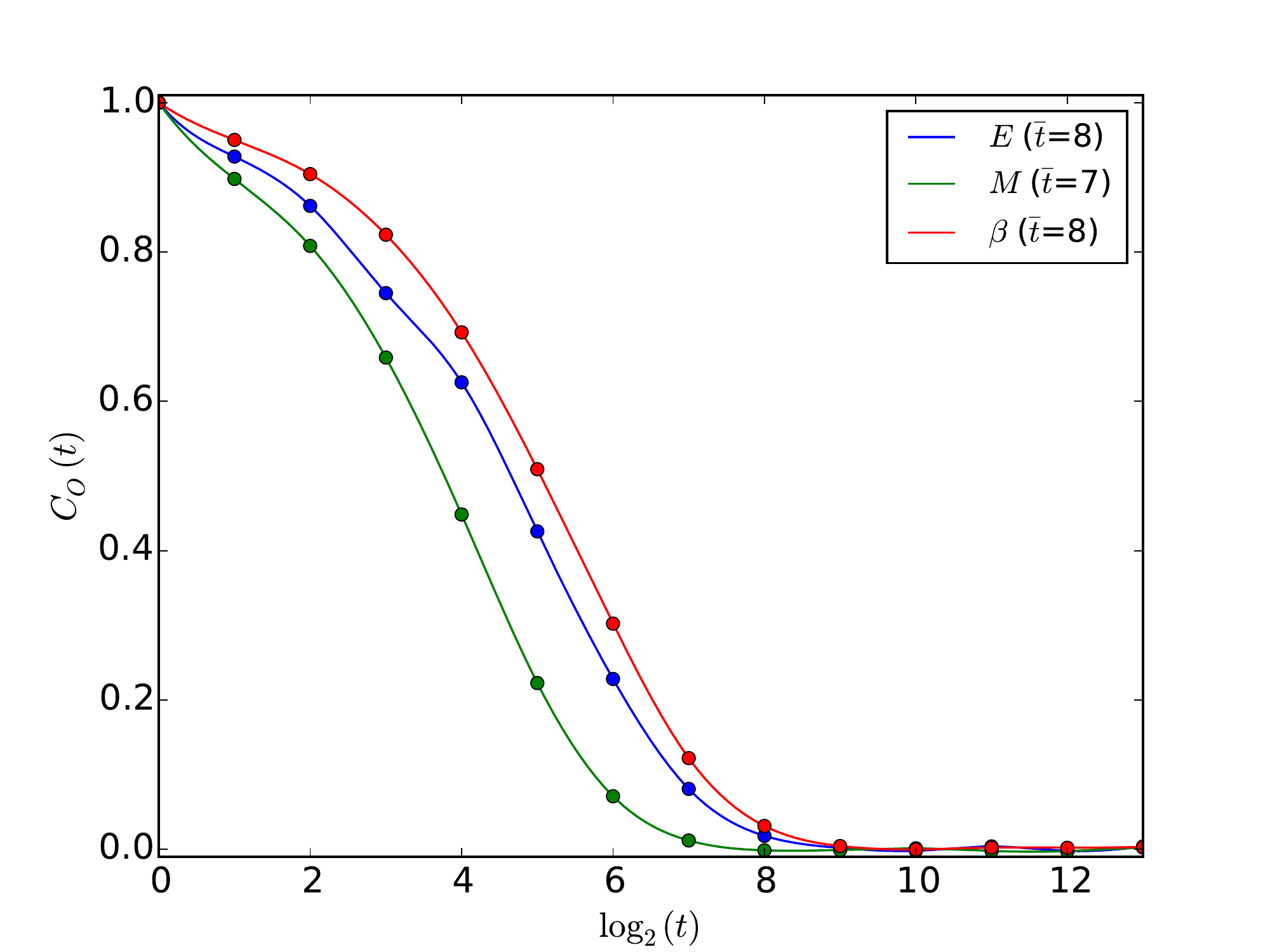}
\includegraphics[scale=0.45]{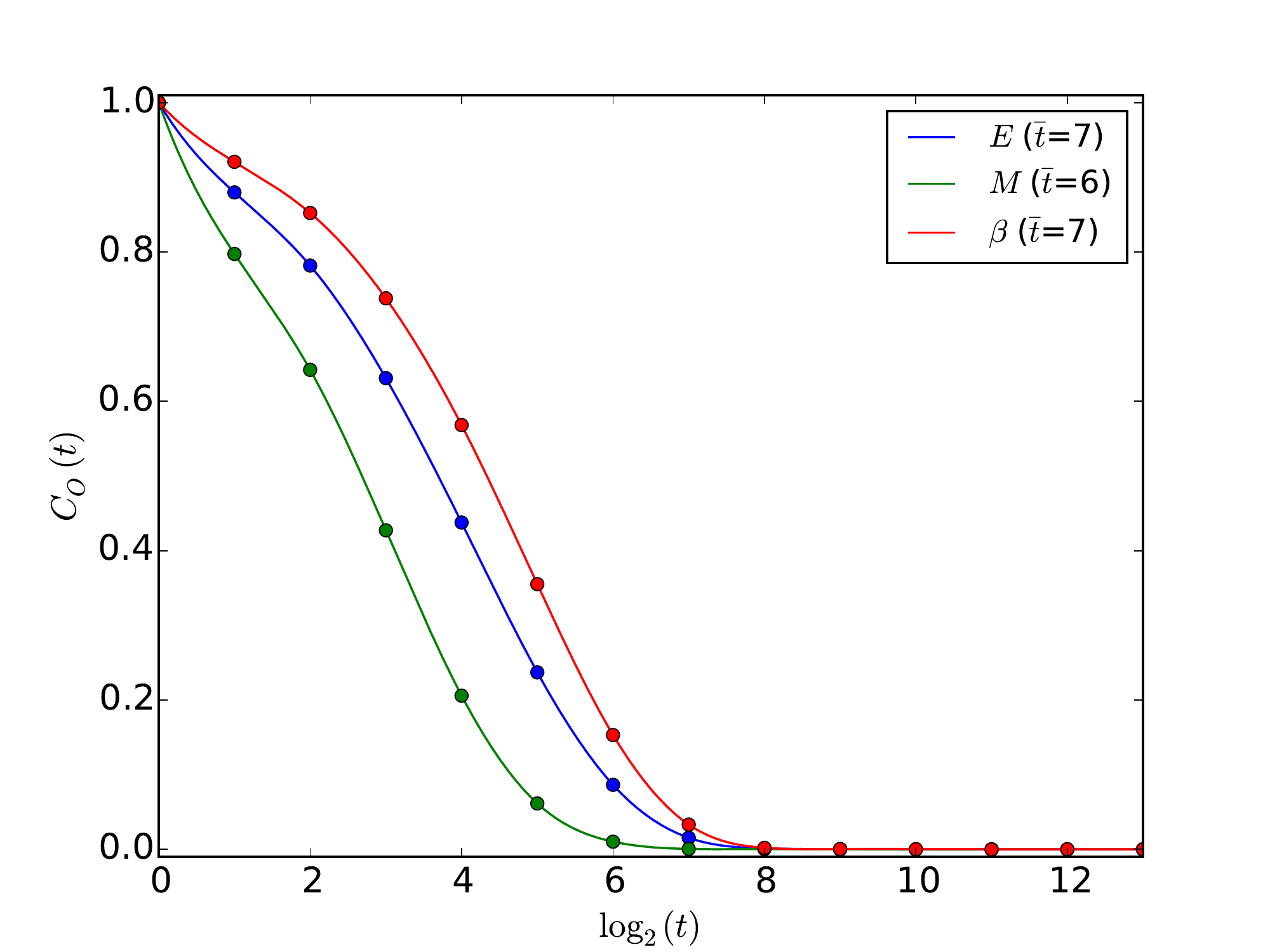}
\caption{(Color online) Autocorrelation function $C_O(t)$ for energy $E$,
magnetization $M$ and inverse temperature $\beta$ as a
function of $\log_2(t)$ for temperature $T$=0.2 (upper panel) and $T$=0.7 (lower panel).}
\label{fig_corr}
\end{figure}

Firstly, the autocorrelation function of the magnetization, energy and inverse
temperature were measured, from which we have obtained an estimation for the decorrelation time.  
For two values of temperature, namely $T$=0.2 and $T$=0.7, we performed longer
simulations, with $n$=8$\times$10$^7$ Monte Carlo sweeps. We have, for these
temperatures, samples of energy, magnetization and $\beta$ which are known
to be correlated because of the intrinsic Markov dynamics implemented in the Monte
Carlo simulation. 
For every observable $O$, in our case the energy $E$, the magnetization $M$, and
the inverse temperature $\beta$, we first computed the autocorrelation function

\begin{equation}
C_O(t)=\frac{\Big<O_i O_{i+t}\Big>-\expect{O}^2}{\expect{O^2}-\expect{O}^2},
\end{equation}
which is plotted as a function of $t$ in Fig. \ref{fig_corr}. We note that, in all cases, 
the correlation becomes negligible for $t \geq 2^{10}$. Also, the estimator of $\beta$ 
takes slightly more time to lose correlation than the other observables.
In this sense, it is a more astringent estimator for statistically independence of the data.

\subsubsection{Binning analysis and central limit theorem}
 
In order to study the statistical properties of the estimator $\hat{\beta}$,
we have performed, as a second independent test, a binning analysis according to
the method outlined for instance, in Refs. \cite{Kawashima1994, Palma2015}, for
temperatures $T$=0.2 and $T$=0.7.

In this method, we divide the sequence of values of an observable $O$ into
blocks of size $k$, so that the total number of blocks is $N_B=\text{int}(n/k)$, where
the \texttt{int} function returns the integer part of its argument. If we denote
the average of the values in the $i$-th block by $\bar{O}_i$, the variance of
these block averages is

\begin{equation}
{\sigma_B}^2(k) = \frac{1}{N_B - 1}\sum_{i=1}^{N_B}(\bar{O}_i - \bar{\bar{O}})^2.
\end{equation}
where $\bar{\bar{O}}$ is the average of all block averages,

\begin{equation}
\bar{\bar{O}} = \frac{1}{N_B}\sum_{i=1}^{N_B} \bar{O}_i.
\end{equation}

\begin{figure}[h!]
\includegraphics[scale=0.45]{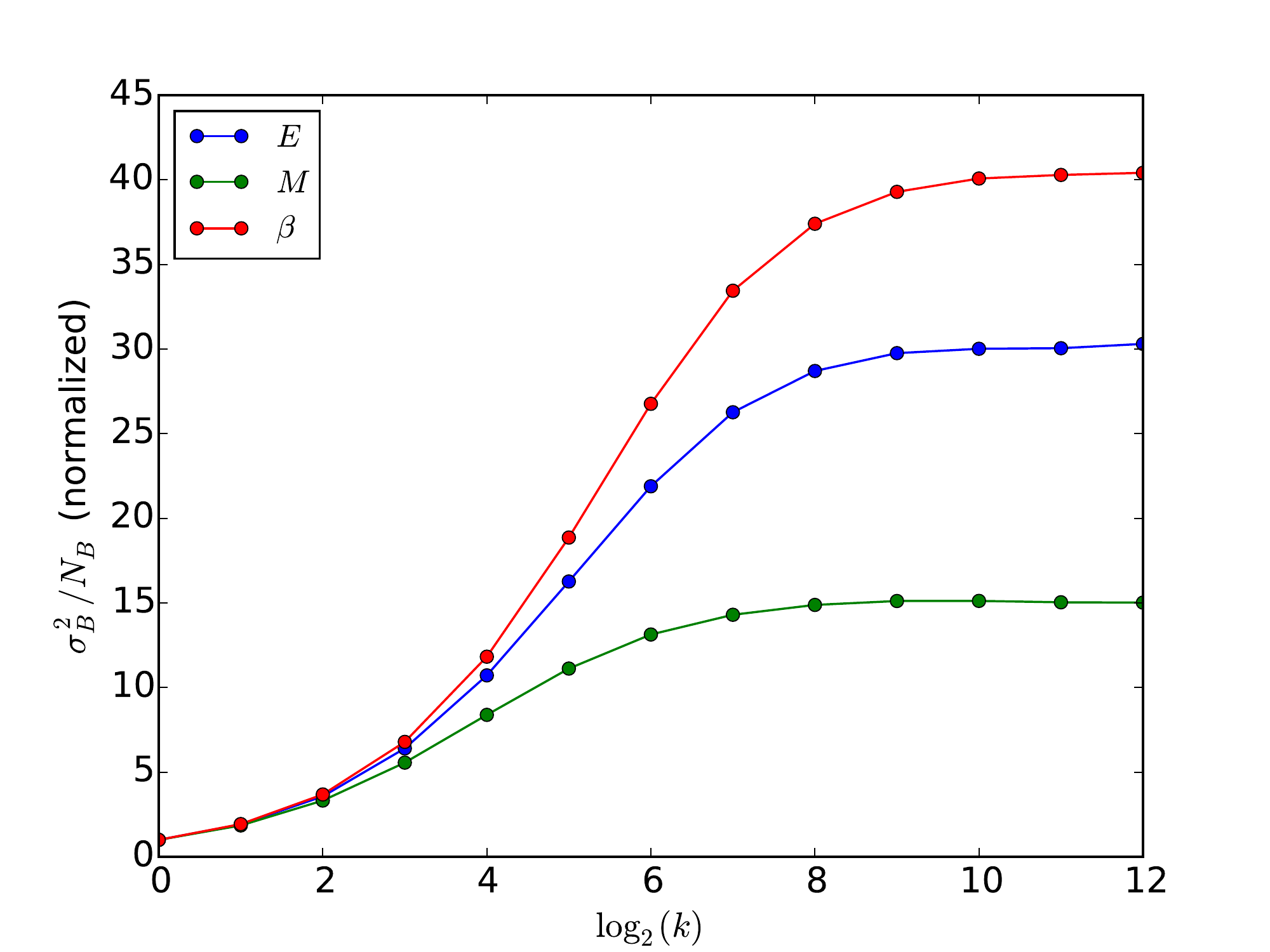}
\includegraphics[scale=0.45]{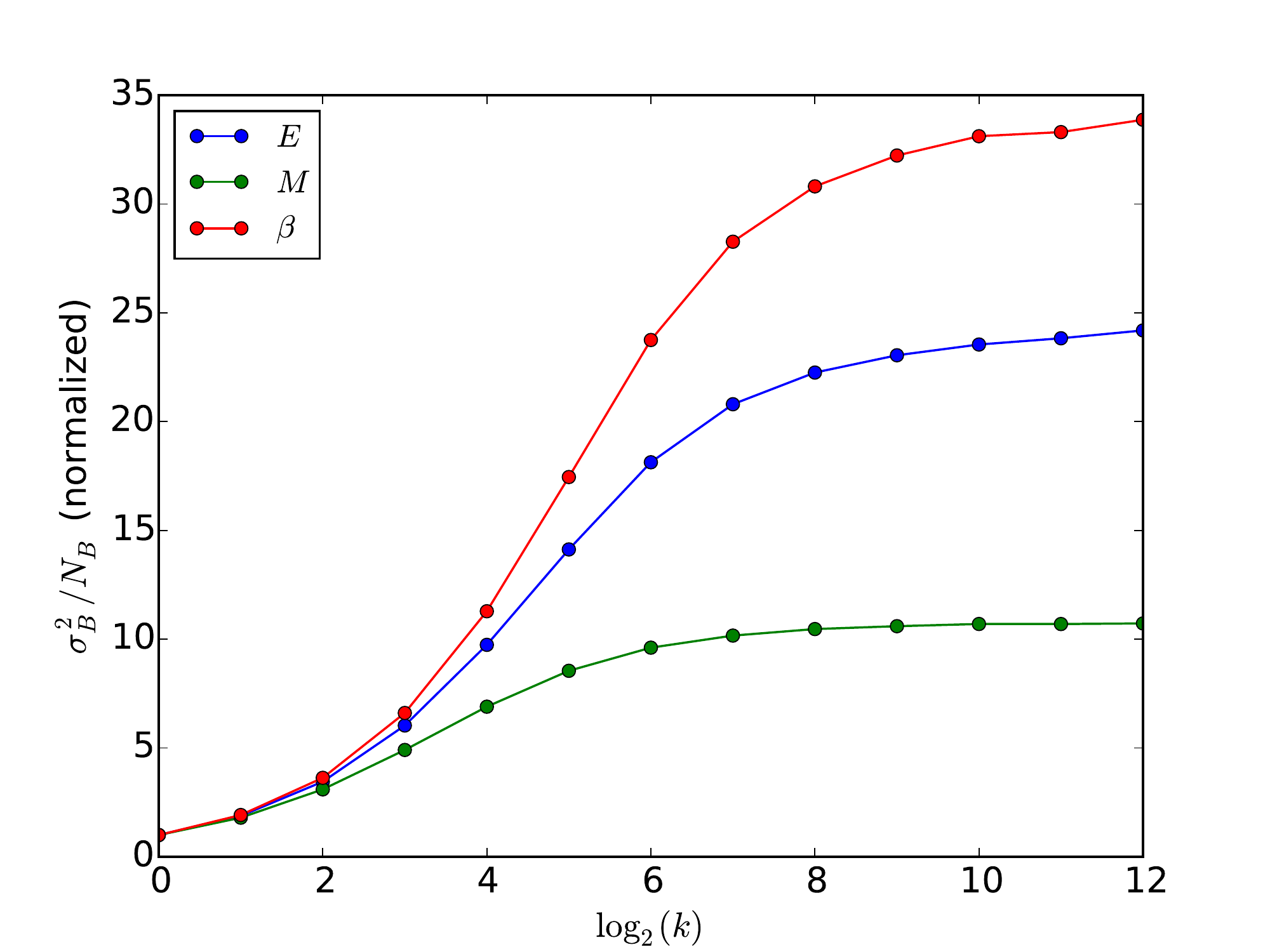}
\caption{(Color online) Binning analysis for energy $E$, magnetization $M$ and
inverse temperature $\beta$, for $T$=0.4 (upper panel) and $T$=0.7 (lower panel).}
\label{fig_binning}
\end{figure}

Under the assumption of statistical independence between the different blocks, the 
variance ${\sigma_B}^2(k)$ should be inversely proportional to $k$, and therefore ${\sigma_B}^2(k)/N_B$ should 
reach a constant value. As we increase $k$, we expect that we approach the
regime where the block averages are really independent from each other. This gives a practical test 
for the minimal block size $k$ that achieves statistical independence. Fig.
\ref{fig_binning} shows this analysis for the observables $E$, $M$, and $\beta$. We see that, as we increase $k$, 
around $k$=2$^{13}$=8192 the quantity ${\sigma_B}^2(k)/N_B$ normalized by ${\sigma_B}^2(1)$ 
reaches a \emph{plateau}, which is consistent with a decorrelation time $t \approx$ 2$^{10}$=1024.

\begin{figure}[h]
\begin{center}
\includegraphics[scale=0.45]{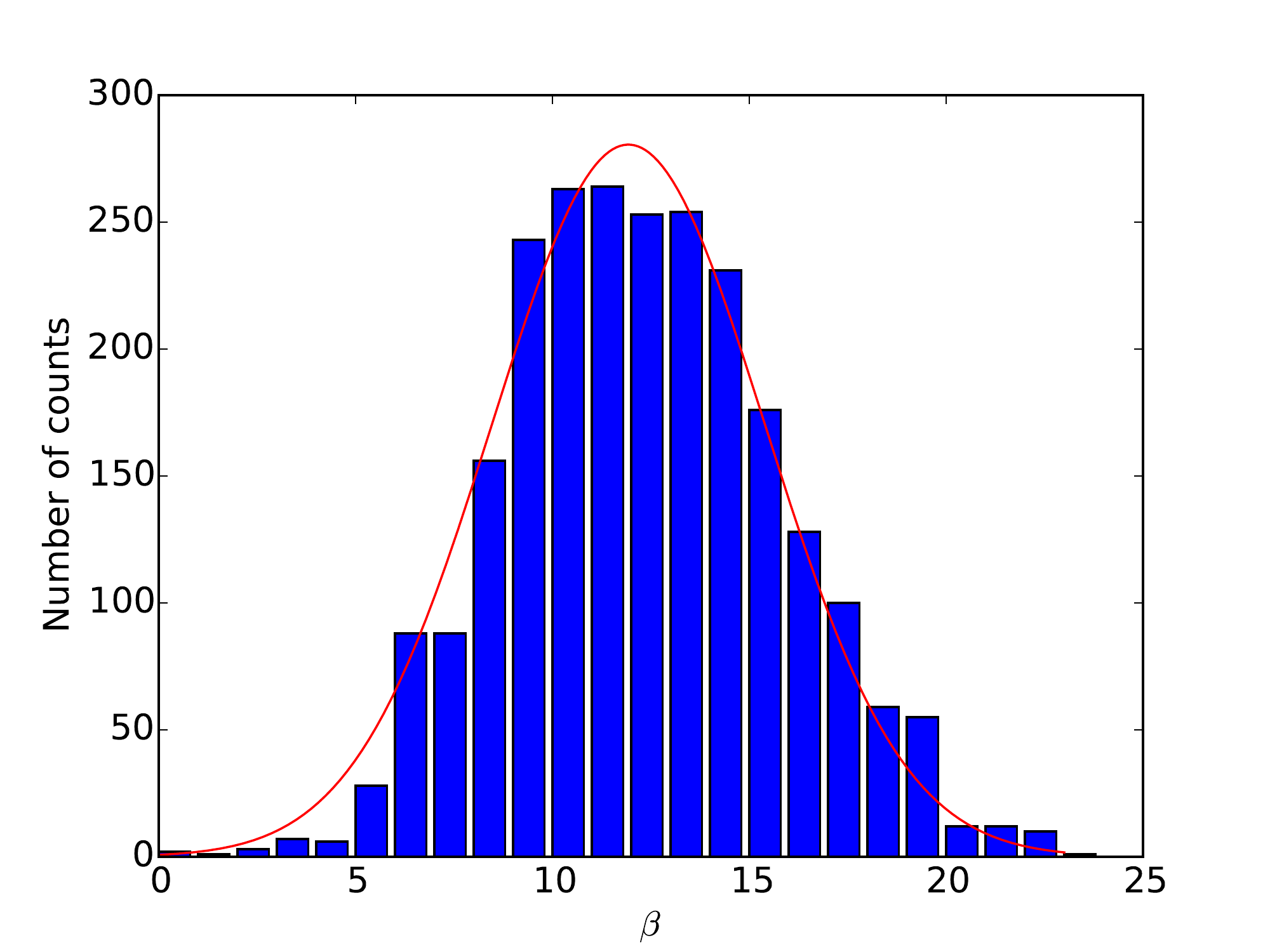}
\includegraphics[scale=0.45]{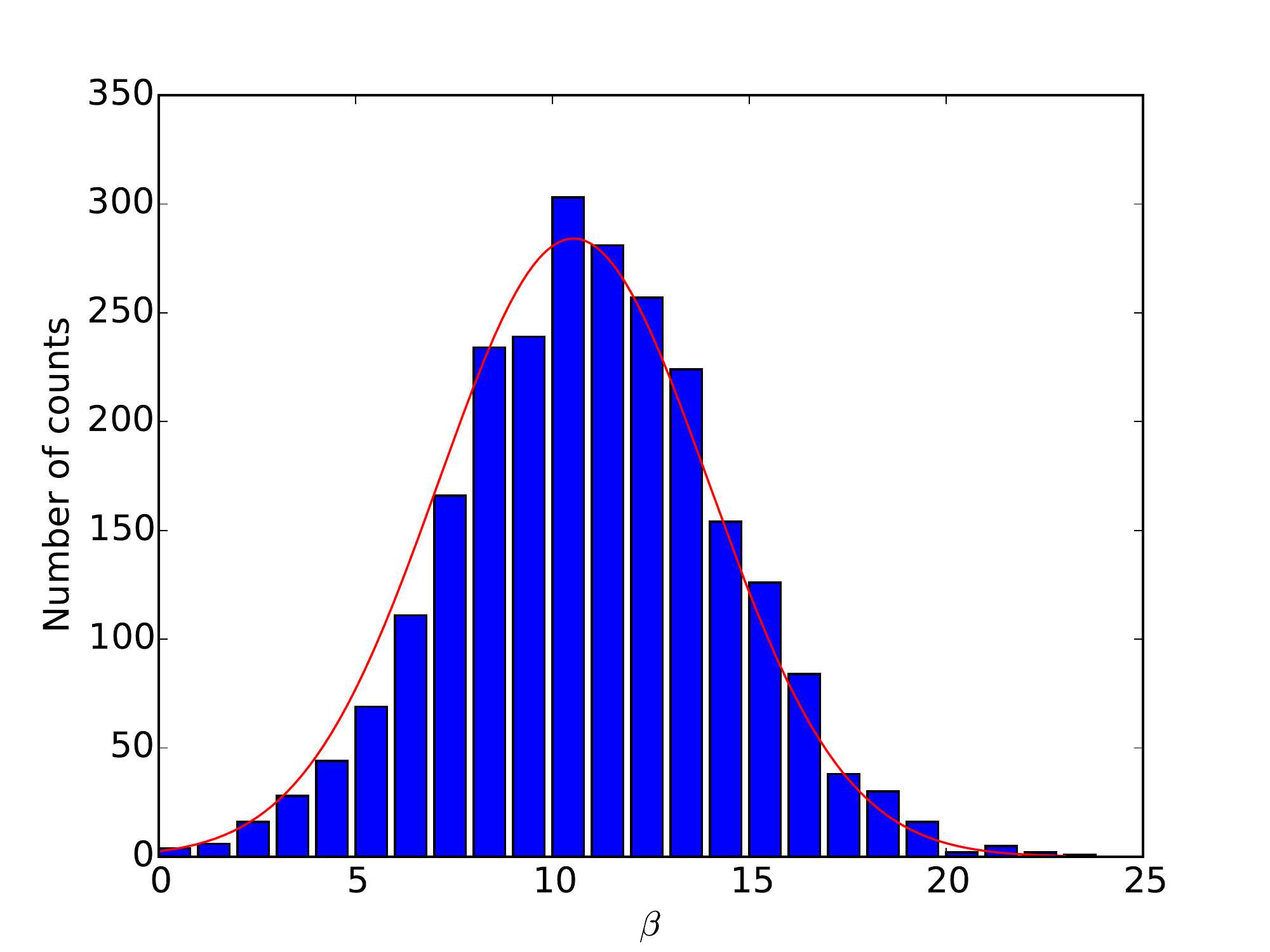}
\end{center}
\caption{(Color online) Histograms of block averages of $\hat{\beta}$ for
$T$=0.4 (upper panel) and $T$=0.7 (lower panel) compared to the corresponding Gaussian
distribution represented by the full, red line.}
\label{fig_histo}
\end{figure}

Finally, to test that the sizes of the thermal averages were large enough to produce independent 
statistics, we have computed the probability density function of the set of values obtained for the average of the magnetization.
We checked the distribution of block averages by constructing histograms of
those averages with block size $k$=2$^{13}$, which are shown in Fig.
\ref{fig_histo}. It can be observed that the histograms approach a Gaussian
distribution as predicted by the central limit theorem.
This criterion gives an estimation for the decorrelation time $\tau$, which is
consistent with the one obtained by the binning analysis.

\section{Conclusions}

In this article we have shown how to construct ensemble-free microscopic
estimators for the inverse temperature. We have demonstrated the
practical usefulness of this estimator by simulating the two dimensional
XY-model in the canonical ensemble. Among other advantages, measuring this
estimator directly as a thermal average over configurations allows to monitor
the transit to equilibrium of the underlying Markov process used in the Monte
Carlo simulation. 

The robustness of the microscopic estimator can be assessed by comparing the
inverse temperature $\beta_I$ and $\beta_M$, resulting in a remarkable agreement
in the whole region of relevant temperatures. The error bars turned
out to be very small, and they represent absolute errors, which give valuable information 
about the efficiency of the algorithm utilized and about the stochastic dynamics.

The idea of constructing ensemble-free microscopic estimators could be extended to other intensive properties such 
as pressure, chemical potential and magnetic field, which may be useful to monitor equilibrium properties of metastable systems.  

\section*{Acknowledgments}
This work was partially supported by Dicyt-USACH Grant No. 041531PA. SD and GG
acknowledge partial funding by CONICYT ACT-1115 and FONDECYT 1140514 (SD).

\bibliography{dyntemp}
\bibliographystyle{unsrt}

\end{document}